\documentclass[aip,jcp,reprint,floatfix]{revtex4-1}

\usepackage{amssymb}
\usepackage{amsmath}
\usepackage{amsfonts}
\usepackage{graphicx}

\newcommand{\fancy}{\mathcal}

\newcommand{\FE}{\fancy{E}}

\newcommand{\lbr}{\left(}
\newcommand{\rbr}{\right)}
\newcommand{\lbrs}{\left[}
\newcommand{\rbrs}{\right]}

\renewcommand{\vec}[1]{\boldsymbol{#1}}

\newcommand{\beq}{\begin{eqnarray}}
\newcommand{\eeq}{\end{eqnarray}}
\renewcommand{\d}{{\rm{d}}}
\newcommand{\tr}{{{\rm{Tr}}}}

\newcommand{\half}{\frac{1}{2}}
\newcommand{\aeq}{\approx}

\newcommand{\rcite}[1]{Ref. \onlinecite{#1}}

\newcommand{\rcites}[1]{Refs \onlinecite{#1}}

\newcommand{\Hx}{\rm{Hx}}

\newcommand{\xrm}{\rm{x}}
\newcommand{\crm}{\rm{c}}
\newcommand{\Hrm}{\rm{H}}
\newcommand{\xc}{\rm{xc}}

\newcommand{\Ec}{E_{\crm}}
\newcommand{\Ex}{E_{\xrm}}

\newcommand{\pr}{^{\prime}}

\newcommand{\vr}{\vec{r}}

\newcommand{\vrp}{\vec{r}\pr}

\newcommand{\vnabla}{\vec{\nabla}}

\newcommand{\bra}[1]{\left\langle#1\right|}
\newcommand{\ket}[1]{\left|#1\right\rangle}

\newcommand{\braketop}[3]{\left\langle#1\left|#2\right|#3\right\rangle}

\newcommand{\ibra}[1]{\langle#1|}
\newcommand{\iket}[1]{|#1\rangle}
\newcommand{\ibraket}[2]{\langle#1|#2\rangle}
\newcommand{\ibraketop}[3]{\langle#1|#2|#3\rangle}
\newcommand{\iee}[1]{\langle#1\rangle_{\FE}}

\newcommand{\orm}{{\rm{o}}}
\newcommand{\Ext}{{\rm{Ext}}}

\newcommand{\EXX}{{\rm{EXX}}}
\newcommand{\LEXX}{{\rm{LEXX}}}
\newcommand{\EEXX}{{\rm{EEXX}}}
\newcommand{\OEXX}{{\overline{\rm{EXX}}}}
\newcommand{\OLEXX}{{\overline{\rm{LEXX}}}}

\newcommand{\hh}{\hat{h}}
\renewcommand{\th}{\hat{t}}
\newcommand{\Vh}{\hat{V}}

\newcommand{\up}{\uparrow}
\newcommand{\down}{\downarrow}
\newcommand{\fup}{f^{\up}}
\newcommand{\fdown}{f^{\down}}
\newcommand{\fsigma}{f^{\sigma}}
\newcommand{\fsigmap}{f^{\sigma'}}

\newcommand{\sigmab}{\bar{\sigma}}

\newcommand{\tsigma}{\theta^{\sigma}}
\newcommand{\tsigmap}{\theta^{\sigma'}}
\newcommand{\tsigmab}{\theta^{\sigmab}}

\newcommand{\rhoh}{\hat{\rho}}

\newcommand{\CU}{C_U}
\newcommand{\CL}{C_L}
\newcommand{\CUt}{\tilde{C}_U}
\newcommand{\Dt}{\tilde{D}}

\newcommand{\hi}{**}

\begin{document}
\title{The flexible nature of exchange, correlation and Hartree physics: %
resolving ``delocalization'' errors in a `correlation free' %
density functional}
\author{Tim Gould}\affiliation{Qld Micro- and Nanotechnology Centre, %
Griffith University, Nathan, Qld 4111, Australia}
\author{John F. Dobson}\affiliation{Qld Micro- and Nanotechnology Centre, %
Griffith University, Nathan, Qld 4111, Australia}
\begin{abstract}
By exploiting freedoms in the definitions of `correlation',
`exchange' and `Hartree' physics in ensemble systems
we better generalise the notion of `exact exchange' (EXX)
to systems with fractional occupations functions of the frontier
orbitals, arising in the dissociation limit of some molecules.
We introduce the Linear EXX (``LEXX'') theory whose pair
distribution and energy are explicitly \emph{piecewise linear} in
the occupations $f^{\sigma}_{i}$. {\hi}We provide explicit
expressions for these functions for frontier $s$ and $p$ shells.
Used in an optimised effective potential (OEP)
approach it yields energies bounded by the
piecewise linear `ensemble EXX' (EEXX) energy and
standard fractional optimised EXX energy:
$E^{\EEXX}\leq E^{\LEXX} \leq E^{\EXX}$.
Analysis of the LEXX explains the success of standard OEP
methods for diatoms at large spacing, and why they can
fail when both spins are allowed to be non-integer so that ``ghost''
Hartree interactions appear between \emph{opposite} spin electrons
in the usual formula. The energy $E^{\LEXX}$ contains
a cancellation term for the spin ghost case. It is evaluated for
H, Li and Na fractional ions with clear derivative discontinuities
for all cases. The $p$-shell form reproduces accurate
correlation-free energies of B-F and Al-Cl.
We further test LEXX plus correlation energy calculations
on fractional ions of C and F
and again shows both derivative discontinuities and good agreement
with exact results.
\end{abstract}
\pacs{31.15.ep,31.15.eg,31.10.+z}
\maketitle

\section{Introduction}

Following initial work by Yang and
coauthors\cite{Yang2000,MoriSanchez2006,Cohen2008,MoriSanchez2009}
on non-interacting ensembles\cite{Perdew1982} with spin-resolved
fractional occupancy, much consideration has been given to the behaviour
of density functional theory (DFT) under the Kohn-Sham (KS)
prescription\cite{HohenbergKohn,*KohnSham}, and its
various common approximations (eg. LDA\cite{KohnSham}, GGA\cite{GGA},
Becke-like\cite{Becke1988}, OEP\cite{OEP1,*OEP2}) in such ensembles.
Many attempts have been made to understand and deal with the issues that
arise in ensembles (see eg. \rcites{Vydrov2007,Cohen2007,Johnson2011}),
with variable success.
We will show that, in such systems, the notion of `correlation' physics
becomes intertwined with `exchange' and `Hartree' physics in
the usual prescription, with (improvable) consequences for
common approximations.

Let us begin by considering, quite generally,
the nature of `electron correlation' and `electron exchange'
in a non-ensemble system. The usual expression for the groundstate
correlation energy can be written as
\begin{equation}
\Ec=
\ibraketop{\Psi}{\hat{H}}{\Psi}-\ibraketop{\Psi^T}{\hat{H}}{\Psi^T}
\label{eqn:Ec}
\end{equation}
where $\hat{H}$ is the Hamiltonian of a many-electron system,
$\iket{\Psi}$ is its groundstate wavefunction, and $\iket{\Psi^T}$ is
some approximation to the wavefunction (by the variational principle,
correlation energy is never positive). Thus correlation
is not an intrinsic property of the system, but a property of the
chosen trial wavefunction. In standard
optimised effective potential (OEP) approaches\cite{OEP1,*OEP2},
including KS DFT, $\iket{\Psi^T}$ takes the form of a single
Hartree-Fock like Slater determinant which is
constructed from one-particle orbitals $\iket{i\sigma}$
evaluated in a \emph{common} one-particle Hamiltonian
$\hat{h}=\th + \Vh$\footnote{we use
atomic units throughout this work such that lengths are in Bohr radii
($1a_0=0.53$\AA) and energies are in Hartree ($1{\rm{Ha}}=4.36$aJ)}
where $\th\equiv-\half\nabla^2$ and $\Vh\equiv V_{\sigma}(\vr)$.
We can now define the exchange energy
$\Ex=\ibraketop{\Psi^T}{\hat{H}}{\Psi^T}-\bar{E}$
and the ``naive Hartree'' energy of the system\footnote{The original
``true Hartree'' theory explicitly excluded orbital
self-interaction, but the ``naive'' form is traditionally used as a
reference in KS DFT}
$\bar{E}=\sum_{i\sigma}\ibraketop{i\sigma}{\th+\Vh_{\Ext}}{i\sigma}
+ \half\int\frac{\d\vr\d\vrp}{|\vr-\vrp|}n(\vr)n(\vrp).$
Here $n(\vr)=\braketop{\Psi^T}{\hat{n}(\vr)}{\Psi^T}
=\sum_{i\sigma}|\phi_{i\sigma}(\vr)|^2$
[where $\hat{n}(\vr)$ is the electron number density operator
and $\phi_{i\sigma}(\vr)=\ibraket{\vr}{i\sigma}$]
and $\Vh_{\Ext}\equiv V_{\Ext}(\vr)$ is the external potential.
The groundstate energy is thus $E=\bar{E}+\Ex+\Ec$ where the
partitioning depends on both the choice of $\bar{E}$ and $\iket{\Psi^T}$.

This can be extended into ensembles by replacing projections on
wavefunctions $O=\ibraketop{\Psi}{\hat{O}}{\Psi}$
by traces on density matrices $O=\tr[\rhoh\hat{O}]$ (where operators
act appropriately for any number of electrons) and
by summing $\bar{E}$ over ensemble members.
The density matrix $\rhoh$ is defined as
\begin{align}
\rhoh=\sum_{\FE}w_{\FE}\iket{\Phi_{\FE}}\ibra{\Phi_{\FE}}
\end{align}
where $0\leq w_{\FE}\leq 1$ is the weight of member $\FE$ with
wavefunction $\iket{\Phi_{\FE}}$ and $\sum_{\FE}w_{\FE}=1$.
Minimisations can then be carried out over $\rhoh$ rather than $\iket{\Phi}$.

\section{Exact exchange approaches}

We can now succinctly define the standard `exact exchange' (EXX)
functional approach. Here we consider only $E^{\EXX}=\bar{E}+\Ex$ with
$\Ec$ assumed to be zero.
Investigations into EXX in fractionally occupied ensemble
systems\cite{MoriSanchez2006,Cohen2009,Makmal2011,Hellgren2012-2}
show both successes and shortcomings (discussed in more detail later).
In all these works, the Hartree and exchange energy takes the
`standard' form, \emph{bilinear} in the occupations $\fsigma_i$:
\begin{align}
E^S_{\Hx}=&\int\frac{\d\vr\d\vrp}{2|\vr-\vrp|}
\sum_{i\sigma j\sigma'}\fsigma_i\fsigmap_j
[P_{i\sigma j\sigma'}-\delta_{\sigma\sigma'}Q_{i\sigma j\sigma'}]
\label{eqn:NEHx}
\end{align}
where $P_{i\sigma j\sigma'}=|\phi_{i\sigma}(\vr)|^2|\phi_{j\sigma'}(\vrp)^2|$
and $Q_{i\sigma j\sigma}=\phi_{i\sigma}(\vr)\phi_{i\sigma}^*(\vrp)
\phi_{j\sigma}^*(\vr)\phi_{j\sigma}(\vrp)$.
Here the negative exchange term cancels
the unphysical positive Hartree interaction of each spin orbital
$\iket{i\sigma}$ with itself. However if two different orbitals
of the same spin are partly occupied
($0<\fsigma_i,\fsigma_j<1$ with $i\neq j$),
or if there is partial occupation of both spins in the same orbital
($0<\fup_i,\fdown_i<1$), there is a
corresponding cross-term in \eqref{eqn:NEHx} that is not cancelled.

In a slightly different context
Gidopoulos \emph{et~al.}\cite{Gidopoulos2002} call this spurious
term the ``ghost interaction'' as it represents an unphysical
interaction between orbitals in
different \emph{non-interacting} ensemble members.
In the regular EXX energy expression \eqref{eqn:NEHx},
the ghost interaction appears in the Hartree and exchange energy terms
involving pairs of orbitals in the frontier orbital.
In a Kohn-Sham interpretation of the equivalent diatom problem,
these interactions would be supressed in the total energy
via orthogonality of the degenerate groundstate wavefunctions.
However, when one does not have the exact exchange-correlation
functional, or as here neglects correlation, it can reappear,
particularly when one does not properly account for the ensemble
nature of the system.

We will argue that, in the ensemble interpretation of
partial occupation\cite{Yang2000,MoriSanchez2006,Cohen2008},
this cross term should not be present,
and its explicit removal results in an improved linear
exact exchange (LEXX) approach which is correctly piecewise
\emph{linear}, not bilinear, in the occupation factors $f$. Here,
defining $\tsigma_{i\FE}$ to be one for orbital $\iket{i\sigma}$
occupied in ensemble member $\FE$ and zero otherwise,
we exploit the fact that the `ensemble occupancy' factor
$\fsigma_i=\iee{\tsigma_i}\equiv\sum_{\FE}w_{\FE}\tsigma_{i\FE}$
requires weights $w_{\FE}$ that are piecewise linear in $\fsigma_i$,
from which it follows that $\iee{\tsigma_i\tsigmap_j}\equiv
\sum_{\FE}w_{\FE}\tsigma_{i\FE}\tsigmap_{j\FE}$
is similarly piecewise linear. All energy terms are proportional
to $\iee{\tsigma_i}$ or $\iee{\tsigma_i\tsigmap_j}$
and are thus piecewise linear. As will be discussed in more
detail later this is equivalent, under an
exchange approach, to finding a non-interacting ensemble of
Slater determinants formed from a \emph{common} set of orbitals
produced in a \emph{common} potential.

This allows the creation of simple functionals that avoid much
of the ``localization and delocalization error'' of
Yang et~al.\cite{Yang2000,MoriSanchez2006,Cohen2008,MoriSanchez2009},
and the ``many electron self interaction error'' of
Perdew et~al.\cite{Perdew2007}.
In the present work we focus on two illustrative cases:
i) a single partially occupied ``frontier'' orbital with
$0\leq \fup_h \leq 1$ and $0\leq \fdown_h \leq 1$; and
ii) open $p$ shells with $\fup_h=\fdown_h$. However the scheme
itself has wider applicability, including the full dissociation
problem of molecules. Ref.~\onlinecite{Gidopoulos2002} might
be considered another specific example of this approach, while
Ref.~\onlinecite{Balawender2005} outlines a similar approach via HF
for the restricted case of fractional occupation of a single spin
(their 1SSO approach).

\subsection{Non-interacting `exchange' ensembles}

To illustrate the general approach we consider, as an example,
ensembles with total and spin-resolved
electron number $N_t=N+f$ and $N_{t\sigma}=N/2+\fsigma$ ($N$ is even).
The groundstate ensemble members and weights can be found be
minimising over density matrices subject to various constraints.
However for simple cases where energy ordering is obvious,
one can construct the ensemble more intuitively, just by demanding
that a given set of occupations $\fsigma_i$ be reproduced.
For example if the frontier orbital is non-degenerate (eg. in an $s$ shell),
then the ensemble will be composed of up to three components.
For $f\leq 1$, the ensemble is formed from $\fup$ parts an $N+1$
electron system with extra electron in $\up$ (short-hand $N+\up$),
$\fdown$ parts $N+\down$ and
$(1-f)$ parts $N$ where, because $N$ is even, both
spins are filled equally. For $f\geq 1$ the ensemble comprises
$(1-\fdown)$ parts $N+\up$, $(1-\fup)$ parts $N+\down$, and
$(f-1)$ parts $N+2$.

The density matrix is composed
of many-electron wavefunctions $\iket{\Phi_{\FE}}$ and is
\begin{align}
\rhoh^{f}=&\sum_{\FE}w_{\FE}\ket{\Phi_{\FE}}\bra{\Phi_{\FE}}.
\label{eqn:PhiEns}
\end{align}
For the present case of a non-degenerate frontier orbial
$w_{\FE}\in \{ 1-f,\fup,\fdown \}$ and
$\Phi_{\FE}\in \{\Phi_{N},\Phi_{N+\up},\Phi_{N+\down}\}$ for $f\leq 1$
while $w_{\FE}\in \{ 1-\fdown,1-\fup,f-1 \}$ and
$\Phi_{\FE}\in \{\Phi_{N+\up},\Phi_{N+\down},\Phi_{N+2}\}$ for $f> 1$.
This leads to a total energy $E(f)=\tr[\rhoh^f\hat{H}]
=\sum_{\FE}w_{\FE}E[\Phi_{\FE}]$ that obeys
\begin{align}
E(f)=&\begin{cases}
f E_{N+1}+ (1-f) E_{N}, & 0\leq f\leq 1
\\
(f-1) E_{N+2} + (2-f) E_{N+1}, & 1<f \leq 2
\end{cases}
\label{eqn:Ef}
\end{align}
where $E_{N}$ is the energy of an $N$-electron system
(note that $E_{N+\up}=E_{N+\down}\equiv E_{N+1}$).

The LEXX is defined in general by assuming
that the trial density matrix $\rhoh^{fT}$ of the ensemble
obeys the same relationship \eqref{eqn:PhiEns} but with the component
wavefunctions $\iket{\Phi_{\FE}}$ replaced by Hartree-Fock like
determinants $\iket{\Phi_{\FE}^T}$
constructed from a \emph{single} set of spin-dependent
orbitals $\{\iket{i\sigma}\}$. This trial density matrix:
i) reduces to the regular EXX for integer occupation, ii) gives
correct energies for H with less than one electron, split arbitarily
between spins, and iii) is constructed from a single
set of orbitals $\ket{i\sigma}$ evaluated in a common Hamiltonian,
a requirement that ensures that OEP or KS methods can be used.
Here the orbitals are
eigen-solutions $\hh\iket{i\sigma}=\epsilon_{i\sigma}\iket{i\sigma}$
of a one-body Hamiltonian $\hh=\th + \Vh$.
We sort the orbitals so that $\epsilon_{i\sigma}\leq\epsilon_{j\sigma}$
for $i<j$.
Taking the spin-resolved density
$n_{\sigma}(\vr)=\tr[\rhoh^{fT}\hat{n}_{\sigma}(\vr)]$
one now finds 
\begin{align}
n_{\sigma}(\vr)=&\sum_i \iee{\tsigma_i}|\phi_{i\sigma}(\vr)|^2
\equiv\sum_i \fsigma_i|\phi_{i\sigma}(\vr)|^2,
\label{eqn:n}
\end{align}
where typically $\fsigma_i=1$ for the inner orbitals
and $\fsigma_h=\fsigma$ where $\ket{h\sigma}$ is the
frontier orbital in the spin-shell with highest energy:
which may or may not be occupied in both spins.

The EXX approximation ($\Ec=0$) allows us to use only
the Hartree and exchange (Hx) components of the pair-density
$n_{2\Hx\sigma\sigma'}\equiv
\tr[\rhoh^{fT}\hat{n}_{\sigma}(\vr)\hat{n}_{\sigma'}(\vrp)]$
to evaluate the electronic groundstate.
From the properties of HF wavefunctions, the pair-density
of an ensemble can be written as
\begin{align}
n_{2\Hx\sigma\sigma'}\equiv&n_{2\Hrm\sigma\sigma'}+n_{2\xrm\sigma\sigma'}
\nonumber\\
=&\sum_{ij}\iee{\tsigma_i\tsigmap_j}
[P_{i\sigma j\sigma'}-\delta_{\sigma\sigma'}Q_{i\sigma j\sigma}].
\label{eqn:n2HxE}
\end{align}
Finally, we can use \eqref{eqn:n2HxE} to calculate the energy
{\hi}via
\begin{align}
E^{\LEXX}=&
\sum_{\sigma}\int\d\vr
\lbrs t_{\sigma}(\vr) +n_{\sigma}(\vr)V^{\Ext}(\vr)
\rbrs
\nonumber\\&
+\half\sum_{\sigma\sigma'}
\int\frac{\d\vr\d\vrp}{|\vr-\vrp|} n_{2\Hx\sigma\sigma'}(\vr,\vrp)
\label{eqn:ELEXXE}
\\
\equiv & \sum_{i\sigma}\iee{\tsigma_i}e^{(1)}_{i\sigma}
+ \sum_{i\sigma j\sigma'}\iee{\tsigma_i\tsigmap_j}e^{(2)}_{i\sigma j\sigma'}
\label{eqn:ELEXXp}
\end{align}
where $t_{\sigma}(\vr)=\sum_i\frac{\iee{\tsigma_i}}{2}
|\vnabla\phi_{i\sigma}(\vr)|^2$ and
\begin{align}
e^{(1)}_{i\sigma}=&\int\d\vr \lbrs \half|\vnabla\phi_{i\sigma}|^2
+ V^{\Ext}|\phi_{i\sigma}|^2 \rbrs,
\label{eqn:ELEXXp1}
\\
e^{(2)}_{i\sigma j\sigma'}=&\half\int\frac{\d\vr\d\vrp}{|\vr-\vrp|}
\lbrs P_{i\sigma j\sigma'}-\delta_{\sigma\sigma'}Q_{i\sigma j\sigma} \rbrs.
\label{eqn:ELEXXp2}
\end{align}
These energy expression are perhaps the most general, and most important
in this work, highlighting the importance of ensemble averages
in the evaluation of average occupation and pair-occupation
factors for groundstate energy calculations.

\subsection{Fractional {$s$} shells}

For the fractionally occupied $s$ shells discussed here,
$\iee{\tsigma_i\tsigmap_j}=\min[\fsigma_i,\fsigmap_j]
-\delta_{ih,jh}\delta_{\sigma\sigmab'}\CU^h$
($\sigmab$ is the opposite spin to $\sigma$ and $\CU^h$ is defined
below). The Hartree and exchange components can be compactly written as
\begin{align}
n_{2\Hrm\sigma\sigma'}=&\sum_{ij} \min[\fsigma_i,\fsigmap_j] P_{i\sigma j\sigma'}
- \delta_{\sigma\sigmab'} \CU^h P_{h\sigma h\sigmab},
\label{eqn:n2H}
\\
n_{2\xrm\sigma\sigma'}=&-\delta_{\sigma\sigma'}
\sum_{ij} \min[\fsigma_i,\fsigma_j] Q_{i\sigma j\sigma}
\label{eqn:n2x}
\end{align}
where we have chosen to split Hartree and exchange
terms via $P$ and $Q$. The term
\begin{align}
\CU^h=\min[\fup,\fdown,(1-\fup),(1-\fdown)]
\label{eqn:CUh}
\end{align}
removes spurious ``ghost interactions'' between electrons
of unlike spin. For a zero to two electron system
equations \eqref{eqn:n2H}-\eqref{eqn:CUh} are equivalent (after integration)
to equation 7 of Ref.~\onlinecite{MoriSanchez2009}
sans the correlation energy term. {\hi}This desirable outcome is
a direct result of the ensemble averaging.

When either $\fup$ or $\fdown$ is integer, $\CU^h=0$
and $n_{2\Hx\sigma\sigma'}\equiv \sum_{ij}\fsigma_i\fsigmap_j
[P_{i\sigma j\sigma'}-\delta_{\sigma\sigma'}Q_{i\sigma j\sigma'}]$ since
$P_{i\sigma i\sigma}=Q_{i\sigma i\sigma}$. Clearly this is the form
used in \eqref{eqn:NEHx} and thus energies derived from
\eqref{eqn:n2H}-\eqref{eqn:n2x} will be identical.
We can now proffer an explanation for the variable success
of the EXX for fractionally occupied ensembles.
By violating the aufbau principle and/or allowing spins to vary in
an unrestricted fashion, good results have been obtained for
atoms and diatoms\cite{Cohen2009,Makmal2011} and systems with fractional
occupancy\cite{MoriSanchez2006}. In these works only one spin was
allowed to be non-integer so that
$\iee{\tsigma_h\tsigmap_h}=\fsigma_h\fsigmap_h$
and the EXX and LEXX energies were equivalent.
In systems where both spins were fractionally occupied
(eg. Refs.~\onlinecite{Cohen2009} and \onlinecite{Hellgren2012-2})
the EXX failed to reproduce the correct derivative discontinuity.
In these works $\fup=\fdown=f/2$ and 
$\iee{\tsigma_h\tsigmap_h}\neq\fsigma_h\fsigmap_h$.
Thus the EXX and LEXX energies differed.
We show later that, in this case, the LEXX is guaranteed to
produce a lower energy.

\subsection{Fractional {$p$} shells}

As a less trivial example, we also consider the case of
degenerate frontier $p$ orbitals with equal densities in each
spin. Here we must sum not only over
ensembles members of different electron number, but also
over the degenerate combinations of $p_x$, $p_y$ and $p_z$
orbitals. For example, in an isolated carbon atom each ensemble
member has fully occupied $1s$ and $2s$ shells, but only
two occupied $2p$ orbitals of the same spin $\sigma$ which we denote
$p_{\gamma}\sigma$ and $p_{\delta}\sigma$ where
$\gamma\neq\delta$ and $\gamma,\delta\in \{x,y,z\}$.
To find the equal-spin, spherically symmetric ensemble
we weight each ensemble equally so that $w_{p_{\gamma}p_{\delta}\sigma}=\frac16$
for all six combinations of $\gamma\neq\delta$ and $\sigma$.
In member $p_{\gamma}p_{\delta}\sigma$
we set $\tsigma_{2p,p_{\gamma}}=\tsigma_{2p,p_{\delta}}=1$
while the remaining $2p$ orbital with spin $\sigma$, and all $2p$
orbitals with spin $\sigmab$ have zero occupation. Averaging
over all cases gives $\iee{\tsigma_{2p,p_{\gamma}}}=\frac13$
as expected, while
$\iee{\tsigma_{2p,p_{\gamma}}\tsigma_{2p,p_{\gamma}}}=\frac13$,
$\iee{\tsigma_{2p,p_{\gamma}}\tsigma_{2p,p_{\delta}}}=\frac16$
for $\gamma\neq\delta$
and $\iee{\tsigma_{2p,p_{\gamma}}\tsigmab_{2p,p_{\delta}}}=0$.

For general unfilled frontier $p$ shells this yields an additional
like-spin correction of the form $-\CL^{h\sigma}[P-Q]$
to \eqref{eqn:n2H} and \eqref{eqn:n2x}
so that
\begin{align}
n_{2\Hrm\sigma\sigma'}=&\sum_{ij} \min[\fsigma_i,\fsigmap_j] P_{i\sigma j\sigma'}
\nonumber\\&
- \lbr\delta_{\sigma\sigma'} \CL^h - \delta_{\sigma\sigmab'} \CU^h \rbr
\sum_h P_{h\sigma h\sigma'},
\label{eqn:n2Hp}
\\
n_{2\xrm\sigma\sigma'}=&-\delta_{\sigma\sigma'}
\sum_{ij} \min[\fsigma_i,\fsigma_j] Q_{i\sigma j\sigma}
\nonumber\\&
- \delta_{\sigma\sigma'} \CL^h \sum_hQ_{h\sigma h\sigmab},
\label{eqn:n2xp}
\end{align}
where we recognise the degeneracy in the outermost $p$ orbitals
by summing over $h$ with equal weights.
Let us restrict ourselves to the case
$\fup=\fdown=f/2$ where $0\leq f<2$ is the total occupation
(over both spins) of each orbital in the shell. One can sum over
the ensemble to show (after much work)
\begin{align}
\CL^{h\sigma}=&\half\min[f,\allowbreak 2-f,\allowbreak |1-f|,\allowbreak 1/3]
\label{eqn:CLh}
\end{align}
for open $p$ shells. We note that the total number of electrons
in the shell is $N_p=3f$ and \eqref{eqn:n2Hp}-\eqref{eqn:CLh} are
valid for $N_p$ integer or fractional.

The like-spin correction ensures that a bilinear approach would
fail even for systems with one spin fully occupied. Indeed it is
only true for the case $\fup=1$, $\fdown=0$ (or vice versa)
occurring for half-occupied shells in N and P. {\hi}Here one must not
only allow the spin-symmetry to be broken, but also break the spherical
symmetry to make the bilinear expression \eqref{eqn:NEHx} correct.

\subsection{General ensemble systems}

{\hi}While we have so far determined our ensembles using explicit
knowledge of the degenerate groundstate, it is possible to
carry out a more general
ensemble minimisation to determine $w_{\FE}$. Here, for
a given potential, we allow ensemble members (determined
by member occupancy factors $\tsigma_{i\FE}$) to sample
all combinations of `occupied' and `unoccupied' orbitals
of the one-electron Hamiltonian,
and minimise the energy with respect to $w_{\FE}$.
In practice we would restrict the allowed ensemble members
to limited combinations predicted to be low in energy.
For example in the $p$ shell case given above, or indeed Be,
we might search for the minimum over cases with full occupancy
in $1s^2$ and varying occupancy in the near-degenerate
$2p$ and $2s$ orbitals.

As shown in equation~\eqref{eqn:ELEXXp},
the general LEXX energy $E^{\LEXX}[\{w_{\FE}\}]$ for a given
ensemble can be written as a sum of the ensemble averaged
occupations $\iee{\tsigma_i}$ and pair-occupations $\iee{\tsigma_i\tsigmap_j}$
with orbital dependent energy prefactors given in
equations~\eqref{eqn:ELEXXp1} and \eqref{eqn:ELEXXp2}.
These averaged occupations depend piecewise linearly on $w_{\FE}$ and thus
$E^{\LEXX}[\{w_{\FE}\}]$ can be minimised under the constraints
$0\leq w_{\FE}\leq 1$ and $\sum_{\FE}w_{\FE}=1$
ie. we look for the (constrained) set of weights minimising
\begin{align}
E^{\LEXX}=&\sum_{\FE}w_{\FE}[
\sum_{i\sigma}\tsigma_{i\FE}e^{(1)}_{i\sigma}
+ \sum_{i\sigma j\sigma'}\tsigma_{i\FE}\tsigmap_{j\FE}e^{(2)}_{i\sigma j\sigma'}
].
\end{align}
We can thus
find, for a given potential and orbtials, the optimal weights
$w_{\FE}$, and through them
$\iee{\tsigma_i}$ and $\iee{\tsigma_i\tsigmap_j}$. For the
true KS potential, this should be equivalent to finding the
temperature$\to 0^+$ limit of finite-temperature DFT.
Such an approach might be useful for dealing with the difficult
atomic dissociation problem.

\section{Optimised effective potentials}

For a many-electron system the EXX (or LEXX) groundstate energy
is composed of the orbital kinetic energy
$T_s=\half\int\d\vr \sum_{i\sigma}\fsigma_i|\nabla\phi_{i\sigma}|^2$,
the energy from the external potential
$E_{\Ext}=\int\d\vr V_{\Ext}n$ and the Hartree plus exchange energy $E_{\Hx}$.
For an ensemble we calculate $E_{\Hx}$ via, for example, the expansion
\eqref{eqn:n2H}-\eqref{eqn:n2x} of $n_{2\Hx\sigma\sigma'}$ for
$s$ shells [or \eqref{eqn:n2Hp}-\eqref{eqn:n2xp} for equi-$p$ shells]
to form the orbital dependent LEXX expression
\begin{align}
  E_{\Hx}=&\sum_{\sigma\sigma'}
  \int\frac{\d\vr\d\vrp}{2|\vr-\vrp|}n_{2\Hx\sigma\sigma'}(\vr,\vrp),
  \label{eqn:EHx}
\end{align}
while for `standard' EXX we instead use \eqref{eqn:NEHx}.
The difference in energies between the LEXX and `standard' EXX
for frontier $s$ shells
is thus the difference between \eqref{eqn:EHx} and \eqref{eqn:NEHx}.
E.g. for the $s$ case
\begin{align}
  E^{\LEXX}-E^{\EXX}=E_{\Hx}-E^S_{\Hx}= - \CUt^he_h
\label{eqn:EfEx}
\end{align}
where $E^S_{\Hx}$ is given by \eqref{eqn:NEHx} and
$e_h=\int\frac{\d\vr\d\vrp}{|\vr-\vrp|}P_{h\up h\down}$
and $\CUt^h=\CU^h-\min[\fup,\fdown]+\fup\fdown
=\min[\fup\fdown,(1-\fup)(1-\fdown)]$
governs the unlike-spin correction to the Hartree energy required
when both $\fup$ and $\fdown$ are non-integer. A similar expression
can be derived for the like-spin correction to $p$-shells.

We can now define orbital dependent groundstate energies via
$E^{\EXX}=T_s+E_{\Ext}+E^S_{\Hx}$ for the EXX and
$E^{\LEXX}=E^{\EXX}-\CUt^he_h$ for the LEXX.
In an optimised-effective potential\cite{OEP1,*OEP2} approach, we look for
a potential $V\equiv V_{\orm\sigma}(\vr)$ such that the orbitals
satisfying $[\th+V_{\orm\sigma}]\phi_{i\sigma}=\epsilon_{i\sigma}\phi_{i\sigma}$
minimise the energy. Here we call this approach the
$\OEXX$ or $\OLEXX$ (with an overline to denote use of an optimised
effective potential) depending on the Hx functional used.
Finding $V_{\orm\sigma}$ involves, as input, the functional derivatives
$D_{i\sigma}(\vr)=\delta E_{\Hx}/\delta\phi_{i\sigma}(\vr)$.
Thus the scheme for finding optimised LEXX solutions
differs only from that for the regular EXX in that
$\Dt_{i\sigma}$ for the LEXX includes an extra term for $i=h$.
Via $\CUt^h$, the additional term vanishes whenever $\fup$ or $\fdown$
is integer, as expected (at least for $s$ shells).

Let us consider some of the formal implications of the LEXX.
Firstly, the total energy found in an optimised LEXX scheme
must be bounded below by the EXX energy of the full ensemble.
To prove this we first note that the ensemble EXX energy $E^{\EEXX}$
for an ensemble of positive weights $w_{\FE}$ of elements $\FE$
can be written as
$E^{\EEXX}(f)=\sum_{\FE} w_{\FE} E^{\EXX}_{\FE}[\{\phi^{\FE}_{i\sigma}\}]$
where $[\th+V^{\FE}_{\orm\sigma}]\phi^{\FE}_{i\sigma}=
\epsilon^{\FE}_{i\sigma}\phi^{\FE}_{i\sigma}$ and
$V^{\FE}_{\orm\sigma}$ is chosen to minimise $E^{\EXX}_{\FE}[\{\phi\}]$
and may vary between different ensemble members.
From \eqref{eqn:n}-\eqref{eqn:EHx}, it is clear that
$E^{\LEXX}[\{\phi_{i\sigma}\}]=\sum_{\FE} w_{\FE} E^{\EXX}_{\FE}[\{\phi_{i\sigma}\}]$
where $V_{\orm\sigma}$ in
$[\th+V_{\orm\sigma}]\phi_{i\sigma}=\epsilon_{i\sigma}\phi_{i\sigma}$
can no longer vary separately for each part of the ensemble.
Thus by the variational nature of an OEP we find
$E^{\EXX}_{\FE}[\{\phi^{\FE}_{i\sigma}\}]\leq E^{\EXX}_{\FE}[\{\phi_{i\sigma}\}]$
and $E^{\EEXX}(f)\leq E^{\OLEXX}$.
Secondly, we see that $E^{\LEXX}[\{\phi\}]\leq E^{\EXX}[\{\phi\}]$
for any set of orbitals $\{\phi\}$ and thus
$E^{\LEXX}[\LEXX]\leq E^{\LEXX}[\EXX]\leq E^{\EXX}[\EXX]$
(where the term in the square brackets labels the OEP
used to evaluate the orbitals)
with the equality holding (for $s$ shells) only when
$\CUt^h=0$ (ie. when each of the spins is integer occupied).
The former inequality follows from
\eqref{eqn:EfEx} by noting that $\CUt^h\geq 0$ and
$e^h=
\int\frac{\d\vr\d\vrp}{|\vr-\vrp|}P_{h\up h\down}\geq 0$
as $P_{h\up h\down}\geq 0$ (similarly for the like spin term) and the latter
follows from the minimisation principle of OEPs.
Putting the OEP inequalities together, we find
\begin{align}
E^{\EEXX}\leq& E^{\OLEXX} \leq E^{\OEXX}
\label{eqn:Ineq}
\end{align}
where we include the overline (indicating an optimised potential was used)
for clarity.

\section{Correlation energies}

The consequences of the improved pair-densities also extends
beyond exchange physics. Some beyond-dRPA correlation energy methods
[see \rcite{Eshuis2012} for an overview]
like the RPAx\cite{RPAx}, RXH\cite{Gould2012-RXH} and PGG\cite{PGG}
kernels, ISTLS\cite{ISTLS,*Gould2012-2}
and tdEXX\cite{Hellgren2008,*Hesselmann2010} depend in some way on the
groundstate pair-density. The difference between the EXX and LEXX
expressions will therefore manifest in \emph{correlation} energies too.
Here we can calculate the correlation energy via the ``ACFD''
functional (see e.g. Ref.~\onlinecite{Eshuis2012})
involving the orbital-dependent linear response function $\chi_0$,
and ``xc kernel'' $f_{\xc}$.
By way of example, the ``PGG''\cite{PGG} kernel directly uses
the pair-density to approximate
\begin{align}
f_{\xc\sigma\sigma'}(\vr,\vrp)\aeq&
\frac{1}{|\vr+\vrp|}
\lbr
\frac{n_{2\Hx\sigma\sigma'}(\vr,\vrp)}{n_{\sigma}(\vr)n_{\sigma'}(\vrp)}
-1 \rbr.
\end{align}
It thus captures
the ensemble physics at both the LEXX and correlation levels via
$n_{2\Hx}$.

\section{Results}

\begin{figure}[thb]
\caption{Groundstate energy differences
$E(\fup,\fdown)-E^{\OLEXX}(\half,\half)$ (Ha)
of H, Li and Na ions with fractional occupations
under EXX (left) and LEXX (right).\label{fig:Enf}}
\begin{tabular}{ll}
\includegraphics[width=0.42\linewidth]{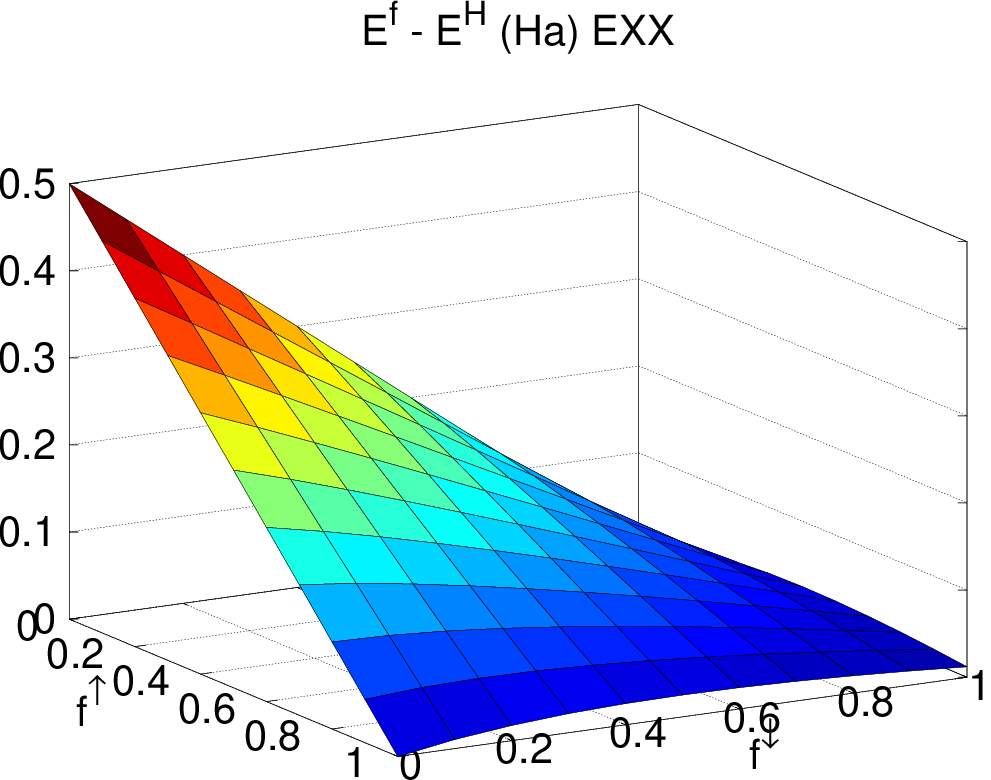}
& \includegraphics[width=0.42\linewidth]{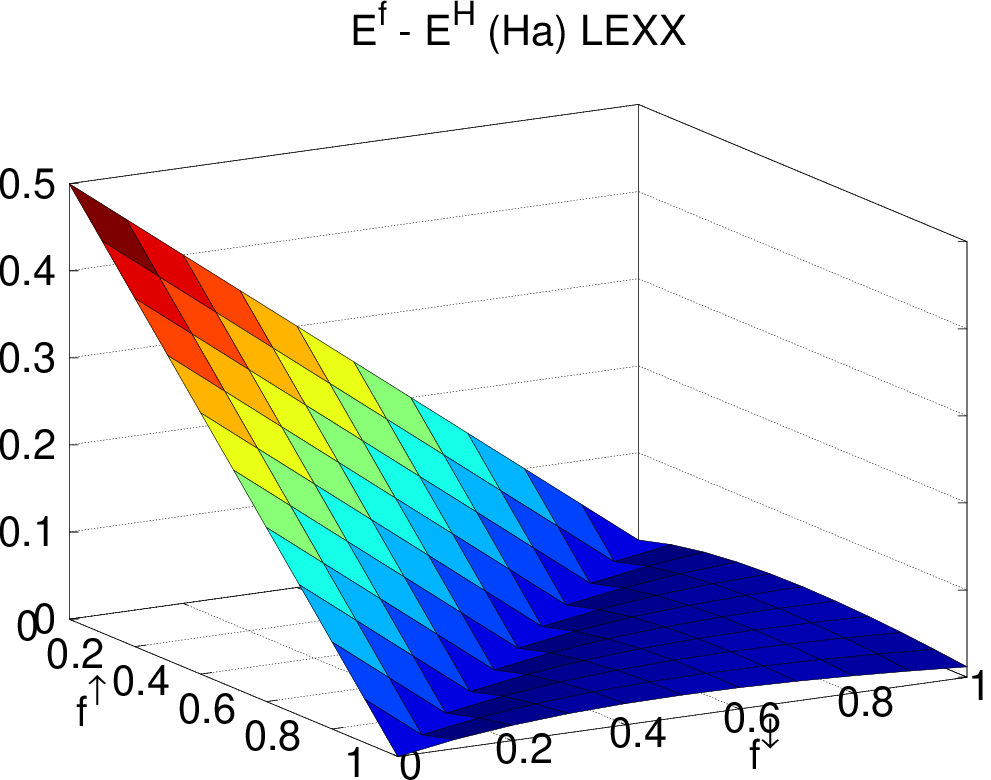}
\\
\includegraphics[width=0.42\linewidth]{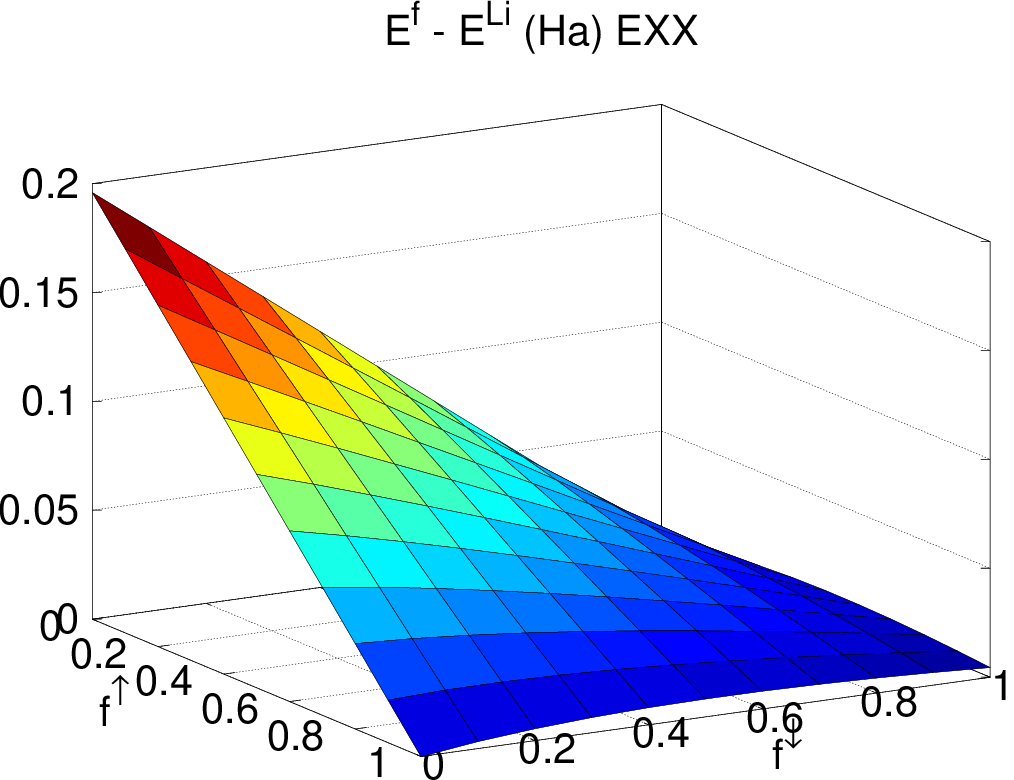}
& \includegraphics[width=0.42\linewidth]{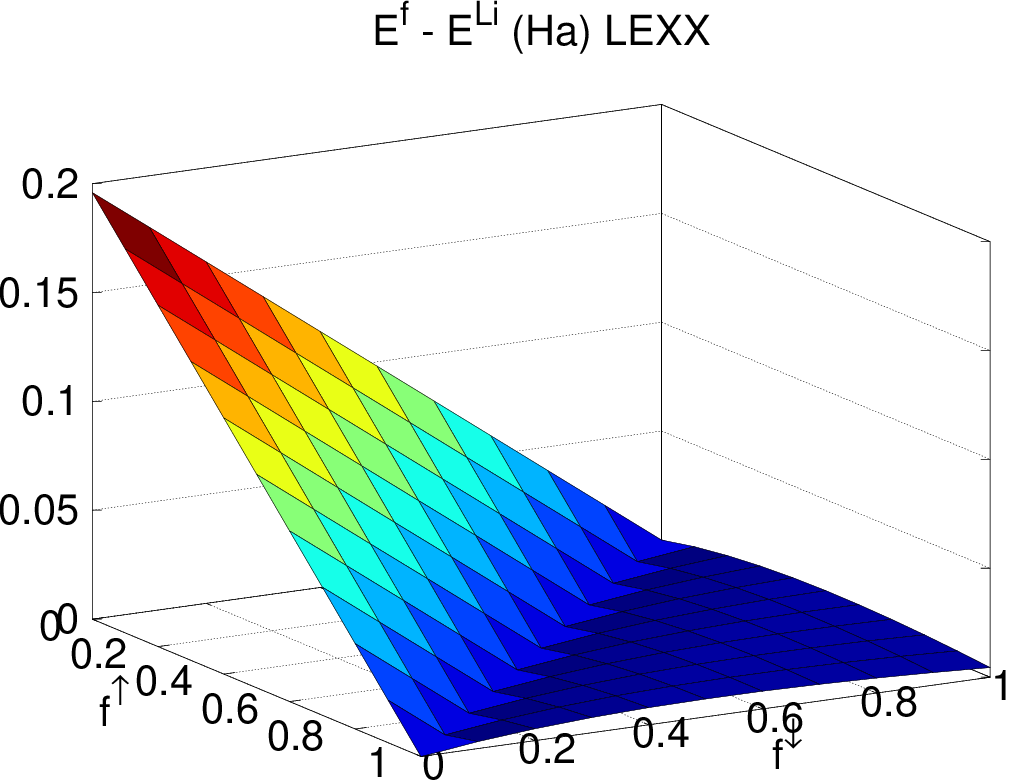}
\\
\includegraphics[width=0.42\linewidth]{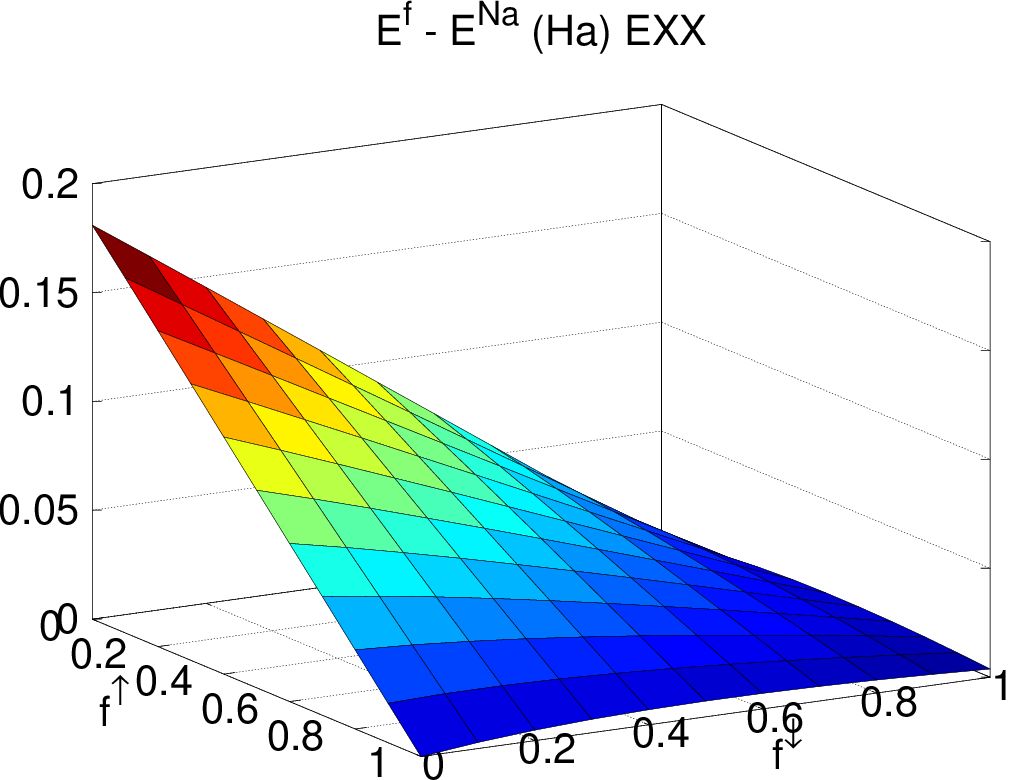}
& \includegraphics[width=0.42\linewidth]{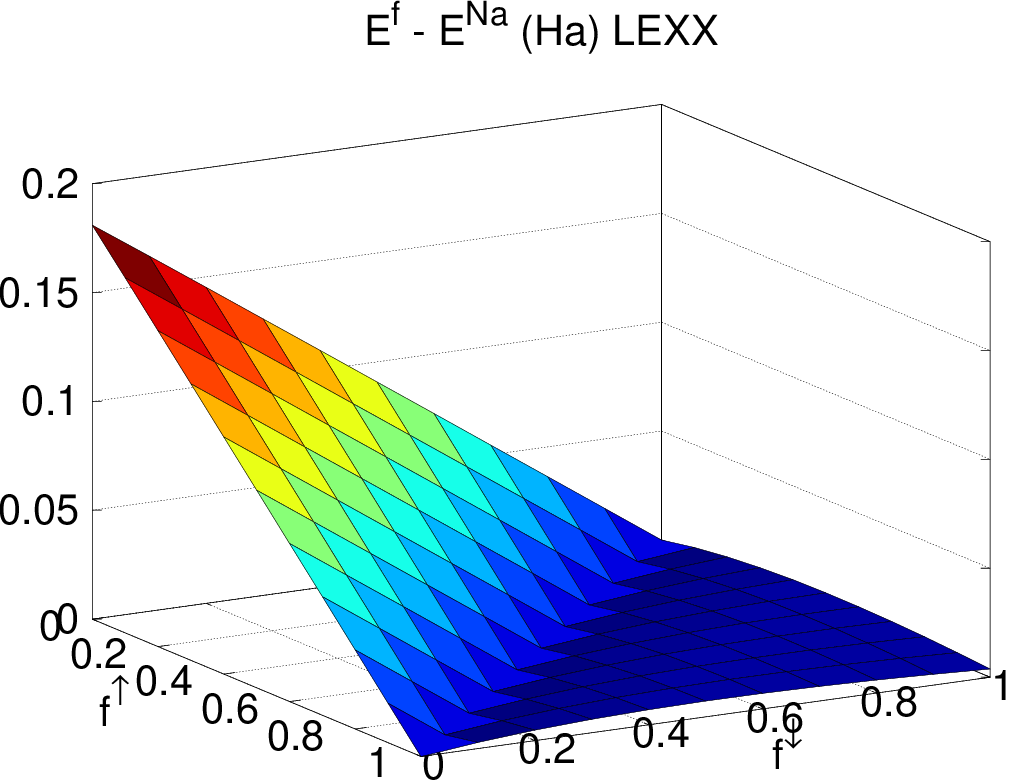}
\\
\end{tabular}
\end{figure}
In Figure~\ref{fig:Enf} we show correlation-free energies for
H, Li and Na-like fractional ions calculated in the optimised EXX and
LEXX schemes under the Krieger, Li and Iafrate\cite{KLI1992} (KLI)
approximation to the potential in a real space code for spherically
symmetric systems. Results are presented for
$\fup$ and $\fdown$ ranging from zero to one 
such that $f$ ranges from zero (e.g. Na${}^+$) to two (e.g. Na${}^-$).
The true ensemble EXX energy $E^{\EEXX}$ takes the same,
piecewise linear form as \eqref{eqn:Ef}
but with groundstate energies $E_N$ of the ensemble members
(for integer $N$) replaced by EXX energies $E_N^{\EXX}$
from the optimal Slater determinant.
The sides of the surface plots show the case where one electron
is integer and the other fractional (or integer at the corners)
and it is clear that the results for the optimised EXX
and LEXX are identical as expected. In the interior,
however, a different picture emerges, with the required derivative
discontinuities at $\fup+\fdown=1$ being absent in the EXX
but clearly present in the LEXX.
The LEXX also varies minimally with $f=\fup+\fdown$ fixed
(along diagonals perpendicular to the projection), unlike the EXX.
The slight remaining non-linearity must be explained via the
implicit dependence
of the orbitals on $f$ as the energy formula is explicitly linear
in $f$. We are unsure if this is a result of the optimised effective
potential approach itself, or the KLI approximation thereto.

The LEXX clearly offers dramatic improvements over the EXX in energy
calculations. For Li and Na it also makes a good approximation
to the true EEXX energy without resorting to correlation physics.
Here the maximum variation from EEXX is at most 6mHa for Li and Na,
significantly smaller than the correlation energies of 45mHa and 396mHa
respectively\cite{Chakravorty1993} for the neutral atoms. Only
for H, where the orbitals of H and H${}^-$ differ significantly through
space, is the difference significant, growing to almost 20mHa
for $f\aeq 1.5$, comparable to the H${}^-$ correlation energy of 42mHa.

{\hi}The LEXX was previously used\cite{Gould2012-RXH} to generate
groundstates for correlation energy calculations. We are thus
able to compare the correlation-free LEXX results from that
work with benchmark HF energies calculated by
Chakravorty \emph{et~al.}\cite{Chakravorty1993}.
To test the validity of the $p$ shell LEXX expression (using
equations~\eqref{eqn:n2Hp}-\eqref{eqn:CLh}) we compared the energies
of the first and second row open $p$ shell atoms B-F and Al-Cl as
these have integer electron numbers, but \emph{fractional} $f$.
For these atoms the LEXX energy has a maximum error of $<1.5$mHa (for O)
and a mean average error of just $0.6$mHa. To numerical accuracy
in our calculations this is close to exact agreement, and
justifies both the LEXX itself and the KLI
approximation to the OEP, at least for integer electron number.

\begin{figure}
\caption{Groundstate energy $E(N)$ of C and F ions
under the LEXX approach with and without correlation
energy included. Electrons are split equally between up and down spin
$N_{\up}=N_{\down}=N/2$.\label{fig:EnC}}
\includegraphics[width=1.00\linewidth]{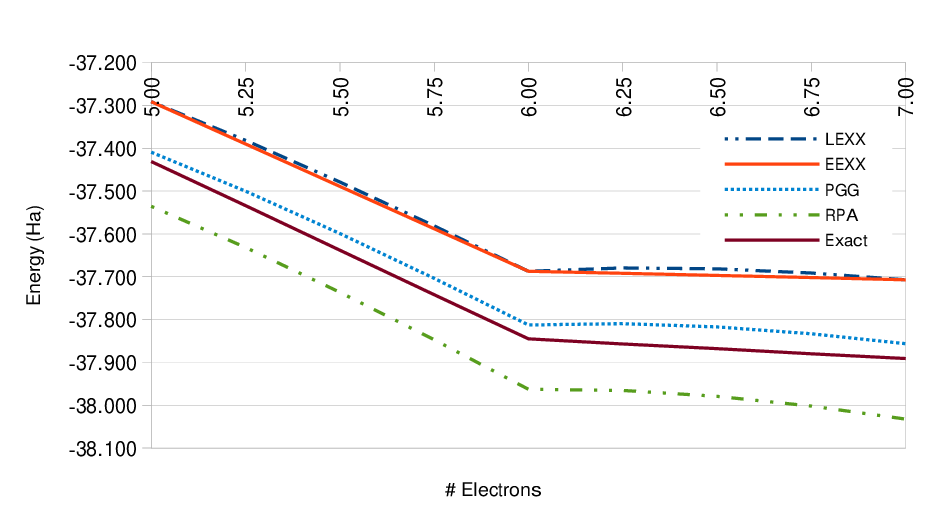}
\\
\includegraphics[width=1.00\linewidth]{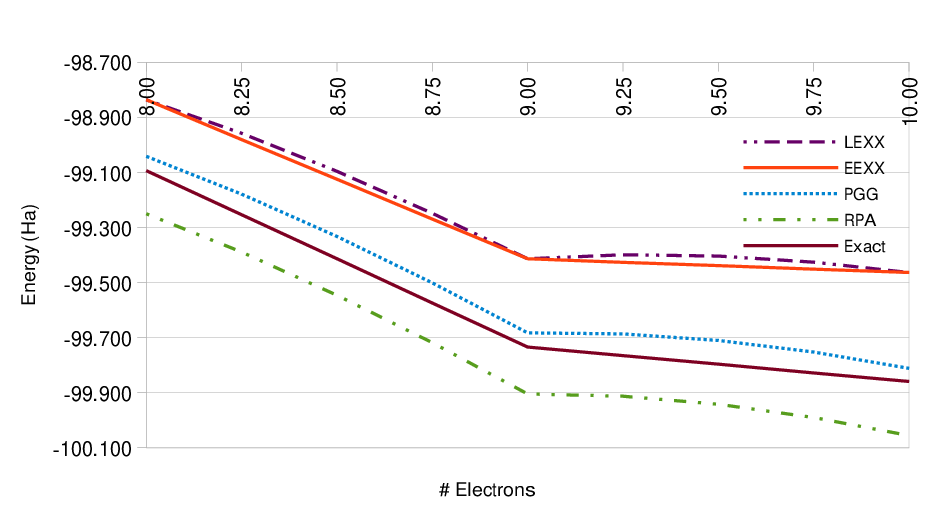}
\end{figure}
In Figure~\ref{fig:EnC}
we show the energy of carbon and fluorine ions with
five/eight to seven/ten electrons. For illustrative purposes
we show results with (RPA, PGG, exact) and
without (LEXX, EEXX) correlation energies
evaluated in the ``ACFD'' functional (see e.g. Ref.~\onlinecite{Eshuis2012}).
The LEXX is used for the kinetic, external, Hartree and exchange
energies in all calculations bar EEXX and exact.
Correlation energies are evaluated using the random-phase approximation (RPA)
and PGG kernel (see Ref.~\onlinecite{Gould2012-RXH} for technical details).
{\hi}The exact groundstate energy of fractional ions of C is given by the
piecewise linear function
$E(N)=E_0^{\rm{C}}-(N-6)I^{\rm{C}}$ for $5\leq N\leq 6$ and
$E(N)=E_0^{\rm{C}}-(N-6)A^{\rm{C}}$ for $6<N\leq 7$ where $E_0^{\rm{C}}$ is the
groundstate energy of carbon, $I^{\rm{C}}$ is its ionisation potential
and $A^{\rm{C}}$ its electron affinity (with similar expression for F).
Energies and ionisation potentials are taken from
Ref.~\onlinecite{Chakravorty1993} and affinities from
Refs.~\onlinecite{Aff4,Aff6}.
The EEXX energy is defined in the same way but with $E_0$, $I$ and $A$
replaced by correlation-free EXX values.

Clearly the LEXX without correlation approximates the
piecewise linear form, albeit
incorrectly predicting negative fractional affinities for
$N\lesssim 6.75$ for C and $N\lesssim 9.60$ for F.
Including correlation improves things, although even here there is
a small range with negative affinities,
at $N\lesssim 6.25$ for C with the RPA and PGG kernels,
and $N\lesssim 9.25$ for F with the PGG kernel. It is clear that
the ``LEXX-PGG'' (PGG evaluated with an LEXX pair-density)
is a fairly good approximation to the groundstate ensemble
energy at all fractions in both cases, especially for the positive ions.
The derivative discontinuity
shown here comes entirely from our correct treatment of Hx in most
cases, with a nonzero but very small extra contribution
from correlation in the PGG case. 
We aim to further investigate
correlation energies at fractional occupation in future work.

\section{Conclusions and further work}

While the discussion here has focused on Fermionic systems with
non-degenerate frontier orbitals and ensembles constructed
around varying electron number, the general approach holds true
for any non-interacting ensemble system.
{\hi}For example in Bosonic systems, orbital SI is not cancelled by
exchange terms even for integer occupation, a situation which
favours the present type of analysis of the ``Hartree'' and ``exchange''
terms. Other interesting cases include finite distance dissociation,
where quantum superpositions of determinants are required as well
as classical ensembles; and thermal ensembles.

LEXX physics is also useful beyond the OEP LEXX method discussed here.
It should be possible to construct
local density functionals (like the LSDA) from pseudo-densities
based on the modified exchange and/or Hartree pair-density
via an approach like that of Ref. \onlinecite{Gidopoulos2012}
or Ref. \onlinecite{Gaiduk2012}.
This perhaps provides some further justification for the success of
recent work by Johnson and Contreras-Garc{\'\i}a\cite{Johnson2011}.
The LEXX may also have potential uses in
$\fancy{O}(N)$-scaling DFT approaches (see \rcite{ONReview}
for a recent review).

By constructing a density matrix with similar properties to the exact
ensemble, we were able to develop an LEXX formalism yielding
an orbital-dependent total energy \eqref{eqn:ELEXXE} via a
pair-density, piecewise \emph{linear} in the occupation factors,
and involving ensemble averages of the one $\iee{\tsigma_i}$
and two $\iee{\tsigma_i\tsigmap_j}$ orbital pair factors
[see \eqref{eqn:ELEXXp}-\eqref{eqn:ELEXXp2}].
This is exemplified for doubly fractional $s$ shells
in \eqref{eqn:n2H}-\eqref{eqn:n2x} with ghost-interactions supressed
by the correction term \eqref{eqn:CUh} and with similar expressions
for $p$ shells discussed in \eqref{eqn:n2Hp}-\eqref{eqn:n2xp} with
additional like-spin correction term \eqref{eqn:CLh}.
Using these energy expressions in the OEP LEXX functional
proposed here gives clearly improved results
(with $E^{\EEXX}\leq E^{\OLEXX}\leq E^{\OEXX}$)
when compared with the more common form of EXX, without resorting
to correlation physics.
This suggests that the very notion of electron correlation is
imprecisely defined for OEP or KS systems with fractional occupancy. 
Using the properties of ensembles to create better trial wavefunctions
and density matrices can be an excellent means of reducing the workload
of the correlation functional in such systems.

\acknowledgments

The authors were supported by ARC Discovery Grant DP1096240. We would
like to thank Maria Hellgren, E. K. U. Gross and J. P. Perdew
for helpful discussion.

\section*{References}
%

\end{document}